\documentclass[]{pasj02} 
\usepackage{comment}
\usepackage{amsmath}
\usepackage{listings}
\usepackage{subcaption}
\usepackage{graphicx}
\usepackage{url}

\usepackage{xcolor}
\newcommand{\commentNO}[1]{{\color{black} #1}} %

\jyear{2025}
\Received{}
\Accepted{}

\begin{document} 

\title{Merger fraction in galaxy groups and clusters at {\boldmath $z <$} 0.2: A non-parametric morphological study with Subaru Hyper Suprime-Cam}

\author{
 Anri \textsc{Yanagawa},\altaffilmark{1}\altemailmark\orcid{0009-0009-3388-2509} \email{yaa\_yanagawa@cc.nara-wu.ac.jp} 
 Yoshiki \textsc{Toba},\altaffilmark{2,1,3,4,5}\orcid{0000-0002-3531-7863}
 Naomi \textsc{Ota},\altaffilmark{1} \orcid{0000-0002-2784-3652}
 Masayuki \textsc{Tanaka},\altaffilmark{3,6} \orcid{0000-0002-5011-5178}
 Nobuhiro \textsc{Okabe},\altaffilmark{7,8,9}\orcid{0000-0003-2898-0728} 
 Ikuyuki \textsc{Mitsuishi},\altaffilmark{10}\orcid{0000-0002-9901-233X} 
 Masatoshi \textsc{Imanishi},\altaffilmark{3}\orcid{0000-0001-6186-8792} 
 Rhythm \textsc{Shimakawa},\altaffilmark{11,12}\orcid{0000-0003-4442-2750}
 Ji Hoon \textsc{Kim}\altaffilmark{13}\orcid{0000-0002-1418-3309} 
 and
 Tomotsugu
 \textsc{Goto}\altaffilmark{14}\orcid{0000-0002-6821-8669}
}
\altaffiltext{1}{Department of Physics, Nara Women's University, Kitauoyanishi-machi, Nara, Nara 630-8506, Japan}
\altaffiltext{2}{Department of Physical Sciences, Ritsumeikan University, 1-1-1 Noji-higashi, Kusatsu, Shiga 525-8577, Japan}
\altaffiltext{3}{National Astronomical Observatory of Japan, 2-21-1 Osawa, Mitaka, Tokyo 181-8588, Japan}
\altaffiltext{4}{Academia Sinica Institute of Astronomy and Astrophysics, 11F of Astronomy-Mathematics Building, AS/NTU, No.1, Section 4, Roosevelt Road, Taipei 10617, Taiwan}
\altaffiltext{5}{Research Center for Space and Cosmic Evolution, Ehime University, 2-5 Bunkyo-cho, Matsuyama, Ehime 790-8577, Japan}
\altaffiltext{6}{Department of Astronomical Science, The Graduate University for Advanced Studies, SOKENDAI, 2-21-1 Osawa, Mitaka, Tokyo, 181-8588, Japan}
\altaffiltext{7}{Department of Physical Science, Hiroshima University, 1-3-1 Kagamiyama, Higashi-Hiroshima,Hiroshima 739-8526, Japan}
\altaffiltext{8}{Hiroshima Astrophysical Science Center, Hiroshima University, 1-3-1 Kagamiyama, Higashi-Hiroshima, Hiroshima 739-8526, Japan}
\altaffiltext{9}{Core Research for Energetic Universe, Hiroshima University, 1-3-1, Kagamiyama, Higashi-Hiroshima, Hiroshima 739-8526, Japan}
\altaffiltext{10}{Graduate School of Science, Nagoya University, Furocho, Chikusa-ku, Nagoya, Aichi 464-8602, Japan}
\altaffiltext{11}{Waseda Institute for Advanced Study (WIAS), Waseda University, 1-21-1, Nishi-Waseda, Shinjuku, Tokyo 169-0051, Japan}
\altaffiltext{12}{Center for Data Science, Waseda University, 1-6-1, Nishi-Waseda, Shinjuku, Tokyo 169-0051, Japan}
\altaffiltext{13}{SNU Astronomy Research Center, Seoul National University, 1 Gwanak-ro, Gwanak-gu, Seoul 08826, Republic of Korea}
\altaffiltext{14}{Institute of Astronomy, National Tsing Hua University, 101, Section2, Kuang-FuRoad, Hsinchu 30013, Taiwan}



\maketitle

\begin{abstract}
We investigate the environmental dependence of galaxy mergers using high-resolution imaging data from the Hyper Suprime-Cam (HSC) Subaru Strategic Program. 
We focus on galaxy groups and clusters at $z < 0.2$ identified by the Sloan Digital Sky Survey as a laboratory of galaxy environment.
We develop a new non-parametric classification scheme that combines the Gini-$M_{20}$ statistics with the shape asymmetry parameter, enabling robust identification of mergers with both central concentration and outer morphological disturbances. 
Applying this method to a sample of 33,320 galaxies at $0.075 \leq z < 0.2$ taken by the HSC, we identify 12,666 mergers, corresponding to a merger fraction of 38\%. 
Our results are consistent with visual classifications from the GALAXY CRUISE project, validating the effectiveness of our method. 
We find that the merger fraction increases with redshift for all subsamples (field galaxies, galaxy pairs, and cluster members), and also shows a strong radial gradient within clusters, increasing toward the center. 
These trends suggest that merger activity is enhanced both at earlier cosmic times and in denser environments, particularly in galaxy groups. We also find tentative evidence that mergers may contribute to AGN triggering in cluster cores. Our study highlights the utility of combining non-parametric morphological diagnostics for large-scale merger identification and provides new insights into the role of environment in galaxy evolution.

\end{abstract}

\section{Introduction}
\label{sec:intro}
Galaxy mergers are considered a crucial aspect for comprehending the evolution of galaxies (e.g., \cite{Conselice14}, and references therein). These mergers often initiate star formation (SF) and/or active galactic nucleus (AGN) activity by increasing the inflow of material on a galactic scale into the vicinity of the nuclear region. Recent studies suggest that the occurrence of galaxy mergers may also be influenced by density environments (e.g., \cite{Lihwai,McIntosh,Perez,Darg,Ellison10,Alonso,Shibuya22,Omori,Laishram,Pearson,Sureshkumar,Puskas}) although whether lower- or higher-density environments are critical for galaxy mergers is a matter of debate (see table 6 in \cite{shibuya25}).

To understand galaxy mergers and their environmental dependencies observationally, it is important to classify mergers accurately.
Various methods have been used to classify mergers: (i) visual classification that is often collaborated with citizen science (e.g., \cite{Lintott,Kartaltepe,Holincheck,Walmsley22,GC_Tanaka,Walmsley}),
(ii) parametric methods such as measuring S\'ersic index \citep{de_Vaucouleurs,Sersic,Aceves}, and (iii) non-parametric methods such as concentration--asymmetry--smoothness (CAS) parameter \citep{Abraham94,Takamiya,Bershady,Conselice03}, Gini--$M_{\rm 20}$ classification \citep{Abraham03,Lotz04,Lotz08b}, the multimode--intensity--deviation (MID) statistics \citep{Freeman13,Peth}, and the shape asymmetry ($A_{\rm shape}$: \cite{Pawlik}). 
More recently, machine-learning (ML) based merger identification \citep{Dominguez,Ciprijanovi20,Ciprijanovi21,Omori,Aussel,Ferreira,Rose} has also been actively utilized, and such methods can predict even the merger stage \citep{Bottrell,Yu-Yen}.

\commentNO{To address environmental dependence of galaxy merger and its connection to AGN activity, the merger fraction ($f_{\rm merger}$) is a crucial parameter. Although many authors investigated how the merger (or galaxy pair) fraction depends on the environment (e.g., \cite{Jian,Duan,Dalmasso,Puskas}), their conclusions remain controversial.

It is now well established that $f_{\rm merger}$ rises with redshift out to $z\sim1-2$ for field galaxies (e.g., \cite{Conselice03}; \cite{Lotz08b,Lotz11}); in quasar-host samples, $f_{\rm merger}$ shows a weak dependence on AGN luminosity \citep{Tang23}.  While several studies find that $f_{\rm merger}$ increases with redshift in massive field galaxies \citep{Lihwai,Man,Eunbin}, other works focusing on lower-mass systems or intermediate-density environments report a decline in $f_{\rm merger}$ at lower redshifts ($z\approx0.03-0.2$; \cite{Nevin}),  and some find no significant redshift trend at all in visually-classified AGN‐host samples (covering $z=0-2$; \cite{Villforth}).  This low-redshift and environment-dependent discrepancy may stem from differences in image quality (e.g., \cite{Bickley24a}), in the definition of environment (e.g., local density vs. halo mass), and in the adoption of ML-based classification methods (e.g., \cite{Margalef}).}

In this work, we revisit the environmental dependence of galaxy merger and its connection to AGN activity in the local Universe ($z < 0.2$) using Hyper Suprime-Cam (HSC: \cite{Miyazaki}), which is mounted on the Subaru Telescope.
We particularly focus on galaxy groups and clusters as a laboratory of galaxy environment.
The high image quality of HSC enables us to find tidal features \citep{Kado-Fong}, dual quasars \citep{Silverman}, and to reveal AGN host properties through the 2D-image decomposition \citep{Junyao,Toba22a}.
For instance, \citet{GC_Tanaka} demonstrated that a non-negligible fraction of galaxies classified as non-merger based on the SDSS image shows merging features on the HSC image.
We optimize a methodology for galaxy merger selection based on Gini--$M_{20}$ and $A_{\rm S}$ calibrated by GALAXY CRUISE, a citizen science project \citep{GC_Tanaka}.

The remainder of this paper is structured as follows.
We describe sample selection and the basic information of the sample in section \ref{sec:data}.
The morphological classification method optimized for galaxy mergers and evaluation of classification accuracy are presented in section \ref{sec:analysis}.
The obtained merger fraction in galaxy groups and clusters, and its dependence on cluster redshift and distance from cluster center are presented in section \ref{sec:result}.
We then discuss the environmental dependence of mergers and its connection to AGN activity in section \ref{sec:discus}.
We summarize the results of this work in section \ref{sec:summary}.
This work assumes a flat Universe with $H_0$ = 67.8 km s$^{-1}$ Mpc$^{-1}$, $\Omega_M$ = 0.308, and $\Omega_\Lambda$ = 0.692 \citep{Planck15}, which are the same as those adopted in \citet{Tempel17}.


\section{Data}
\label{sec:data}
\subsection{Sample selection}
\label{subsec:selection}
\begin{figure}[h]
 \begin{center}
  \includegraphics[scale=0.4]{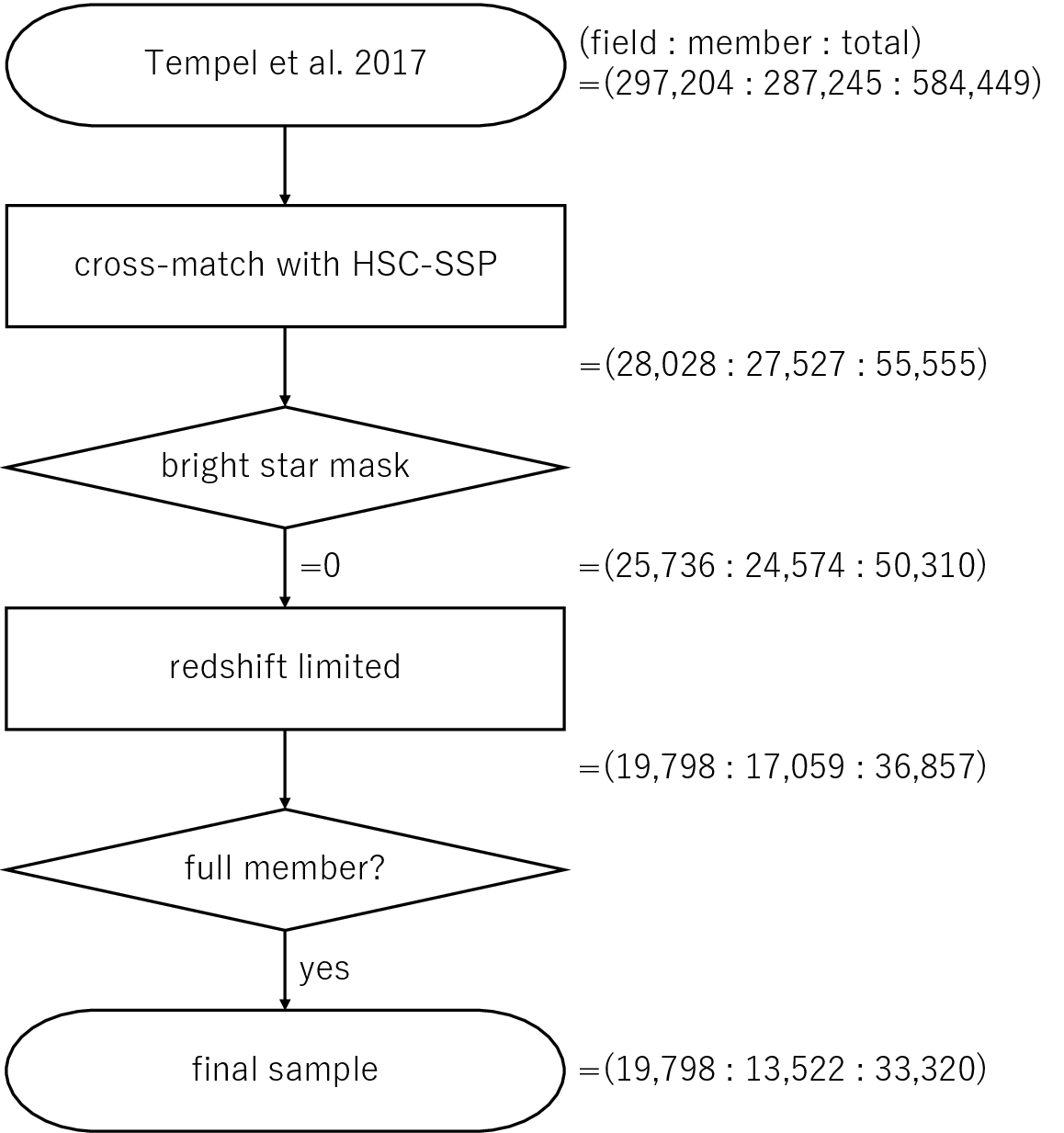}
 \end{center}
\caption{Flowchart of the sample selection process. Each step of the selection pipeline is illustrated, along with the number of galaxies retained at each stage. The counts are presented separately for field galaxies, member galaxies in galaxy clusters and groups, and their total.
{Alt text: flowchart depicting the selection process for galaxy samples used in image analysis. The chart begins at the top with an initial sample of 584,449 galaxies, consisting of 297,204 field galaxies and 287,245 member galaxies. The flow proceeds downward through a series of filtering steps, each applying specific selection criteria.}
}
\label{fig:flow}
\end{figure}

A summary of the sample selection process and the number of galaxies retained at each stage is shown in figure~\ref{fig:flow}.
We constructed our galaxy sample based on the group and cluster catalog of \citet{Tempel17}, which identifies galaxy associations using the Friends-of-Friends (FoF) algorithm in the Sloan Digital Sky Survey (SDSS; \cite{York}) spectroscopic sample . 
The availability of spectroscopic redshift information in SDSS enables us to assume a high probability of physical association among galaxies classified as members.

To obtain high-resolution morphological data, we cross-matched this catalog with HSC Subaru Strategic Program (HSC-SSP; \cite{Aihara18a}, \yearcite{Aihara18b}, \yearcite{Aihara19}, \yearcite{Aihara22})\footnote{We used {\tt s21a\_wide} data.}
The HSC-SSP is an optical-imaging survey covering approximately 1200~$\rm deg^2$ with five broadband filters and approximately 30~$\rm deg^2$ with five broadband and four narrowband filters (see \cite{Bosch,Coupon,Furusawa,Huang,Kawanomoto,Komiyama})\footnote{See also \citet{Schlafly}, \citet{Tonry}, \citet{Magnier}, \citet{Chambers}, \citet{Juric}, and \citet{Ivezic} for relevant papers.}.
While the SDSS images have a typical seeing of $\sim1.1$ arcsec \citep{Ross}, the HSC offers significantly finer seeing of 0.6 arcsec. 
By focusing on galaxies present in both SDSS and the HSC-SSP region, we are able to utilize high-quality HSC imaging with associated redshift information, which is essential for accurate merger classification.

The cross-matching between the SDSS-based catalog and HSC data was performed with a positional tolerance of 1 arcsec. 
To mitigate contamination from bright nearby sources, we applied selection filters based on the bright-star mask \citep{Coupon} provided in the HSC catalog (see also section 4.2 in \cite{Aihara22}). 
The mask flags potential contamination based on the proximity and brightness of neighboring stars. 
In our analysis, we excluded objects flagged with \texttt{\_blooming}, \texttt{\_ghost}, or \texttt{\_halo} in the $i$-band. 
We did not apply the broader \texttt{\_any} mask or the \texttt{\_dip} flag because the \texttt{\_dip} condition tends to be triggered more frequently for non-merger galaxies with low interaction probability $P(\mathrm{interaction})$ in the GALAXY CRUISE sample (see section~\ref{ssubsec:GC}). 
Applying it would artificially lower the number of non-mergers and inflate the merger fraction.

We also excluded galaxies with redshifts below $z = 0.075$ to avoid classification inaccuracies caused by excessive spatial resolution (see section~\ref{ssubsec:accuracy} for details). Additionally, for galaxies belonging to groups or clusters, we retained only those systems for which all member galaxies had corresponding HSC imaging data, ensuring completeness. This criterion led us to exclude clusters and groups that are close to the edge of the HSC-SSP survey footprint. After applying these criteria, our final sample consisted of 33,320 galaxies, including both field galaxies and group/cluster members. 

\subsection{Sample properties}
\label{subsec:sample_prop}
\begin{figure}
 \begin{center}
  \includegraphics[scale=0.27]{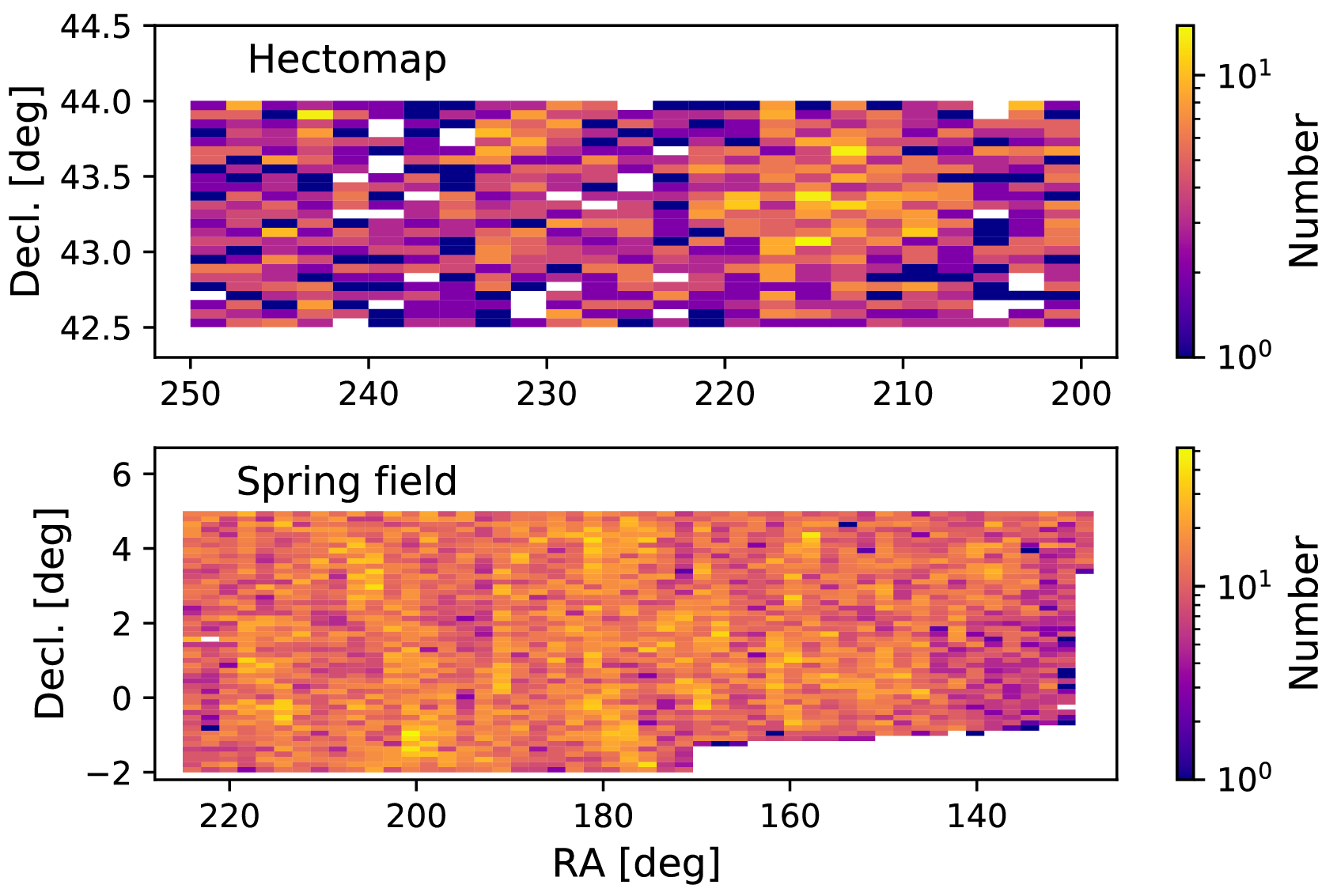}
 \end{center}
\caption{Sky distribution of the selected galaxies in the HSC-SSP region. The color scale represents the number density of galaxies per pixel.
{Alt text: Two two-dimensional histograms showing the sky distribution of the selected galaxy samples. Hectmap region and Spring region from HSC-SSP are plotted.}
}
\label{fig:sky}
\end{figure}

\begin{figure}
 \begin{center}
  \includegraphics[width=0.4\textwidth]{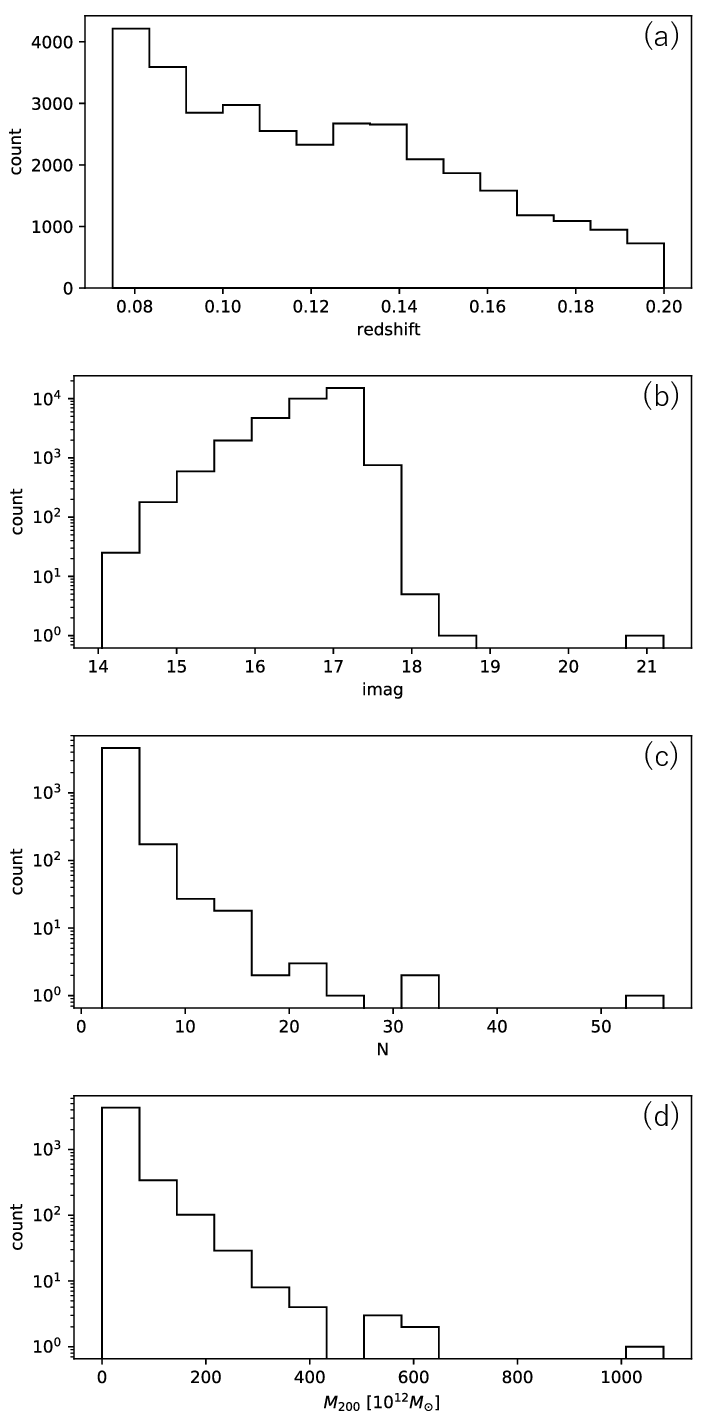}
 \end{center}
 \caption{
 The histograms of sample properties: the distributions of (a) redshift, (b) $i$-band magnitude, (c) richness ($N$), and (d) cluster mass ($M_{200}$), where (a) and (b) represent all galaxies, and (c) and (d) correspond to the 4,827 galaxy groups and clusters. 
{Alt text: Four histograms are vertically aligned. (a) to (d) in order from top to bottom.}
 }
 \label{fig:hist}
\end{figure}

The final sample consists of 33,320 galaxies located within the HSC-SSP survey region, as shown in figure~\ref{fig:sky}. 
These galaxies have spectroscopic redshifts in the range $0.075 \leq z < 0.2$, and their $i$-band magnitudes span from 14.0 to 21.2. 
The redshift and $i$-band magnitude histograms are illustrated in figures~\ref{fig:hist}a and \ref{fig:hist}b.

We also confirmed that 4,827 galaxy groups and clusters are located in the HSC-SSP survey footprint.
To characterize the galaxy environments, we also examine the histograms of galaxy richness ($N$, defined as the number of galaxies in a given group or cluster) and cluster mass ($M_{200}$, defined as the mass enclosed within a radius where the mean density is 200 times the critical density of the Universe), as presented in figures~\ref{fig:hist}c and \ref{fig:hist}d. 
It is worth noting that the majority of our sample resides in group-scale systems with richness $N < 18$.

\section{Analysis}
\label{sec:analysis}
\subsection{Calculating morphological diagnostics}
\label{subsec:statmorph}
To classify galaxy morphologies, we adopted a non-parametric approach by computing structural indicators directly from imaging data. 
For this purpose, we used the Python package \texttt{statmorph}\footnote{We used v0.5.7 (\url{https://github.com/vrodgom/statmorph/releases/tag/v0.5.7}).} (\cite{RGV2019}, \yearcite{RGV2022}), which is designed to calculate a range of morphological diagnostics for galaxies.

\texttt{statmorph} is primarily based on the methodology introduced by \citet{Lotz04}, along with subsequent improvements implemented in the IDL codes (\cite{Lotz06}, \yearcite{Lotz08a}, \yearcite{Lotz08b}).
Given a galaxy cutout image, a corresponding weight map, and the point spread function (PSF), the code computes several diagnostic quantities, including the Gini--$M_{20}$ statistics \citep{Lotz04}, the CAS parameters \citep{Conselice03}, and the MID statistics \citep{Freeman13}.
It also performs two-dimensional S\'{e}rsic model fitting.

\begin{figure*}
 \begin{center}
  \includegraphics[width=0.9\textwidth]{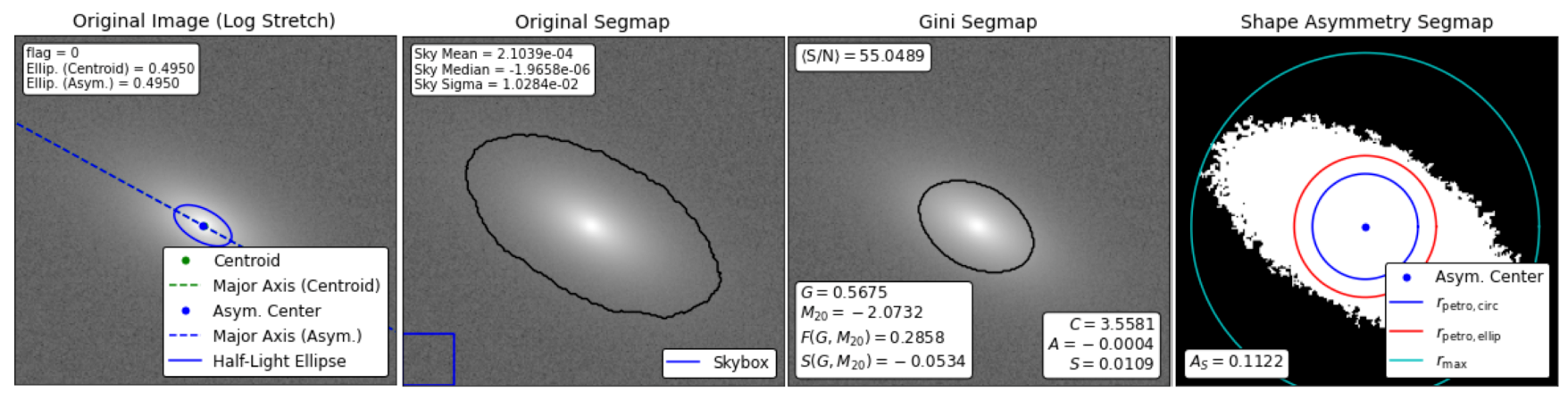}
 \end{center}
\caption{
Example outputs from \texttt{statmorph} applied to an HSC galaxy image. The panels show:
    left: the original galaxy cutout image.
    Second from left: the original image overlaid with the segmentation map identifying galaxy pixels.
    Third from left: the region used for calculating the Gini--$M_{20}$ parameters.
    Right: the detection mask used for computing the \texttt{shape\_asymmetry} statistic.
See \citet{RGV2019} for further details on the calculation procedures and definitions.
{Alt text: One big image with four panels.}
}
\label{fig:ex}
\end{figure*}

In this study, we applied \texttt{statmorph} to the HSC $i$-band images using cutouts with a radius eight times the Petrosian radius\footnote{The Petrosian radius was obtained from the SDSS DR10 \texttt{PhotoObjAll} table, where we adopted the \texttt{petroR90\_i} parameter.}, following the procedure described in \citet{Khanday}. We also extracted the PSF image for each galaxy using the {\tt PSF picker} tool. Among the outputs produced by \texttt{statmorph}, we focused on two key morphological indicators for merger classification: the Gini--$M_{20}$ statistics and the \texttt{shape\_asymmetry} parameter. These parameters were selected for their sensitivity to morphological disturbances, which are characteristic of ongoing or recent mergers.
Figure~\ref{fig:ex} illustrates an example of the outputs by \texttt{statmorph}, including the fitted S\'{e}rsic model, residual images, segmentation maps, and detection masks used in the diagnostic calculations. For further details, see \citet{RGV2019}.

\subsubsection{Gini--$M_{20}$ statistics}
\label{ssubsec:G-M}
The Gini--$M_{20}$ method \citep{Lotz04} provides a quantitative framework for identifying galaxy mergers based on the concentration and distribution of light in a galaxy image. The Gini coefficient ($G$) measures the inequality of the pixel flux distribution, with higher values indicating stronger central concentration. The $M_{20}$ parameter quantifies the second-order moment of the brightest 20\% of the galaxy's light and is sensitive to spatial extent and substructure.

These two quantities are combined into a linear statistic defined as:
\begin{equation}
S(G, M_{20}) = 0.139\,M_{20} + 0.990\,G - 0.327,
\end{equation}
following \citet{Snyder15}. Galaxies with $S(G, M_{20}) > 0$ are considered merger candidates. This method is widely used due to its sensitivity to asymmetric features and multiple light concentrations that often indicate merging activity.

\subsubsection{Shape asymmetry}
\label{ssubsec:shape_asym}
The \texttt{shape\_asymmetry} parameter measures the degree of morphological asymmetry in a galaxy, based on its segmentation map rather than its flux image \citep{Pawlik}. This makes it particularly effective for detecting faint tidal features and extended disturbances in low surface-brightness regions.

The parameter is calculated by rotating the segmentation map 180 degrees around its centroid and comparing it to the original:
\begin{equation}
A_{\mathrm{shape}} = \frac{\sum_{i,j} |I_{ij} - I_{ij}^{180}|}{\sum_{i,j} |I_{ij}|} - A_{\mathrm{bkg}},
\end{equation}
where $I_{ij}$ and $I_{ij}^{180}$ are the original and rotated pixel values, and $A_{\mathrm{bkg}}$ is the background asymmetry correction.
Unlike traditional asymmetry measures, \texttt{shape\_asymmetry} is more sensitive to outer, low-signal morphological disturbances and complements the Gini--$M_{20}$ criteria. If a galaxy has a \texttt{shape\_asymmetry} value greater than 0.2, it is identified as a merger in 95\% of cases with tidal features, according to the benchmark analysis by \citet{Pawlik}.

\subsubsection{A new method for galaxy merger classification}
\label{ssubsec:merger_detect}
To improve merger identification accuracy, we propose a combined classification criterion using both $S(G, M_{20})$ and $A_{\mathrm{shape}}$.
Each of these indicators is sensitive to different morphological signatures of mergers: $S(G, M_{20})$ captures concentrated and asymmetric light distributions, while $A_{\mathrm{shape}}$ excels at detecting faint tidal features and extended disturbances.

We define a galaxy as a merger candidate if it satisfies both of the following conditions:
\begin{equation}\label{eq:bunrui}
S(G, M_{20}) > 0 \quad \text{and} \quad A_{\mathrm{shape}} > 0.2.
\end{equation}
Because both metrics are computed within the segmentation map associated with each galaxy, this dual criterion enables us to detect merger features manifested in either brightness structure or geometric irregularity. The combined use of these indicators enhances robustness, particularly for galaxies with low surface brightness features that might be missed by traditional methods.
This classification scheme forms the foundation for our subsequent analysis of merger fractions and environmental trends. 
\commentNO{The novelty of our method is calibrating combined Gini-$M_{20}$ and $A_{\mathrm{shape}}$ thresholds using the GALAXY CRUISE interaction parameter. Section 3.2.3 and figure 5 demonstrate this yields a higher correlation and improves both completeness and purity compared to single-metric approaches.}

\subsection{Evaluation of classification accuracy using GALAXY CRUISE}
\label{ssubsec:GC}
GALAXY CRUISE is a large-scale citizen science project conducted in the same HSC-SSP region as this study \citep{GC_Tanaka}. Over two million independent visual classifications were collected for 20,686 galaxies with redshifts $z \leq 0.2$. These visual classifications are treated as the ground truth in our evaluation of the \texttt{statmorph}-based merger classification.

\begin{figure*}[h]
\begin{center}
\includegraphics[width=0.9\textwidth]{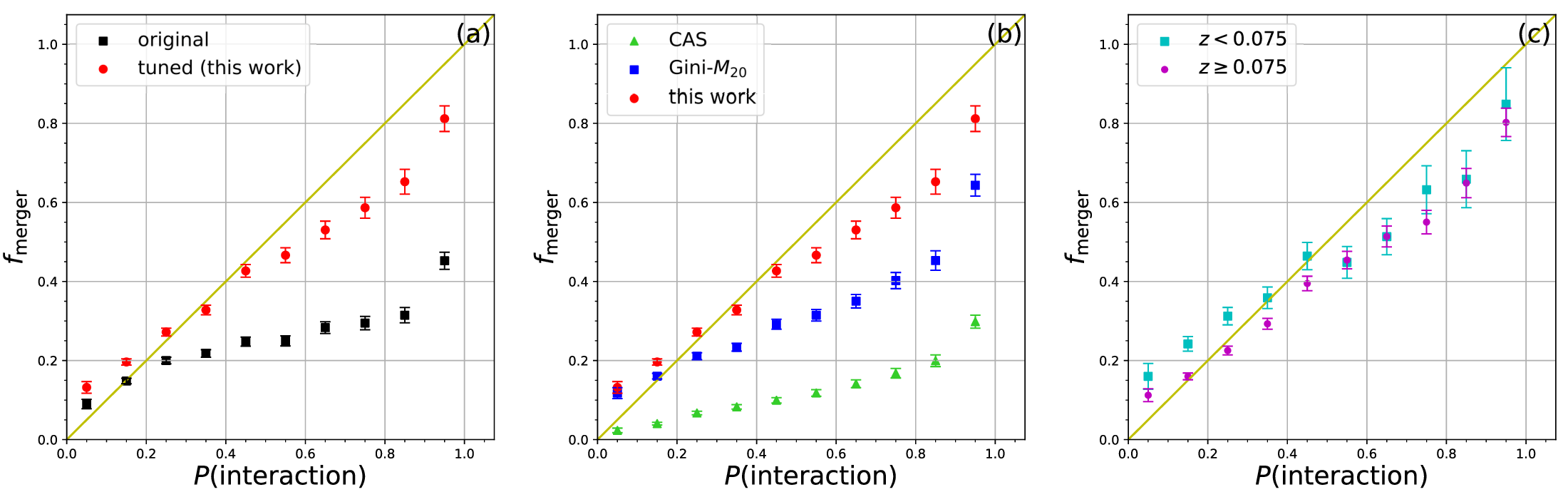}
\end{center}
\caption{
Evaluation of merger classification accuracy using GALAXY CRUISE visual data. 
(a) Merger fraction $f_{\mathrm{merger}}$ as a function of interaction probability $P(\mathrm{interaction})$. Black squares indicate results from the default \texttt{statmorph} settings, while red circles represent results using our adjusted configuration. The yellow solid line denotes the one-to-one relation where $f_{\mathrm{merger}} = P(\mathrm{interaction})$.
(b) Comparison of different classification methods applied to the adjusted sample. Green triangles: CAS statistics; Blue squares: Gini--$M_{20}$ statistics; Red circles: the combined method proposed in this study (equation (\ref{eq:bunrui})).
(c) Same as panel (a), but shown separately for two redshift bins. The analysis reveals a drop in classification accuracy at low redshift ($z < 0.075$), motivating a redshift cutoff.
{Alt text: Graphs of classification accuracy comparisons labeled a to c. The x axes show the $P(\mathrm{interaction})$ from 0.0 to 1.075. The y axes show the merger fraction from 0.0 to 1.075.}
}\label{fig:rate}
\end{figure*}

\subsubsection{Definition of merger fraction}
\label{ssubsec:merger_fraction}

We define the merger fraction $f_{\mathrm{merger}}$ as the proportion of mergers within a given bin:
\begin{equation} \label{eq:f_merger}
f_{\mathrm{merger}} = \frac{N_{\mathrm{merger}}}{N_{\mathrm{total}}}.
\end{equation}
The associated uncertainty is calculated using standard error propagation. 
Assuming Poisson statistics, the error is given by:
\begin{align}
\delta f_{\mathrm{merger}} &= f_{\mathrm{merger}} \sqrt{\frac{1}{N_{\mathrm{merger}}} + \frac{1}{N_{\mathrm{total}}}}.
\end{align}

\subsubsection{Accuracy comparison with visual classification}
\label{ssubsec:accuracy}
Each galaxy in the GALAXY CRUISE sample is assigned two classification probabilities: $P(\mathrm{spiral})$ and $P(\mathrm{interaction})$, both ranging from 0 to 1. A higher $P(\mathrm{interaction})$ indicates a greater likelihood of the galaxy being visually classified as a merger (see \cite{GC_Tanaka}, for more detail). 
We use $P(\mathrm{interaction})$ as the horizontal axis binning parameter and compute the merger fraction in each bin using our automated classification.

Figure~\ref{fig:rate}a compares the results from \texttt{statmorph} using default settings (black) and our adjusted settings (red). The ideal one-to-one relation is indicated by the yellow line. Our adjusted method shows significantly better alignment with the visual classification trend, indicating improved accuracy.
Figure~\ref{fig:rate}b presents a comparison of different classification methods applied to the same dataset. CAS (green), Gini--$M_{20}$ (blue), and our combined method (red, using equation (\ref{eq:bunrui})) are shown. Our method identifies more mergers across a wider range of $P(\mathrm{interaction})$, demonstrating improved sensitivity to faint or subtle merger signatures.

To evaluate redshift dependence, figure~\ref{fig:rate}c shows the same analysis as (a), broken down by redshift bins. 
At lower redshifts ($z < 0.075$), the agreement between $P(\mathrm{interaction})$ and the merger fraction decreases, suggesting an increased rate of false positives. We therefore excluded galaxies with $z < 0.075$ from our final analysis.

\subsubsection{Final thresholds for detection}

Based on the evaluation above, we adopted the following detection threshold in our modified \texttt{statmorph} setup:
\begin{align}
\texttt{cutout\_smooth} \geq \texttt{ellip\_annulus\_mean\_flux} \cdot \frac{0.2}{\texttt{asym} + 0.25}
\end{align}
(see Appendix \ref{ap:1} for further details). 
This empirically tuned criterion improves the robustness of morphological measurements against noise, especially in low surface-brightness galaxies.

\section{Results}
\label{sec:result}
\subsection{Merger classification}
\label{subsec:class}

We applied the merger classification method defined in equation (\ref{eq:bunrui}) to our galaxy sample using the adjusted \texttt{statmorph} configuration.
We note that \texttt{statmorph} outputs a flag that indicates the quality of the basic morphological measurements.
It takes an integer from 0 to 4. 
If it has 2 or more, it means the calculation was bad (see \cite{RGV2019} for further details).
Hence, galaxies with flag $\geq$ 2 were considered ``unclassified'' in this work.

\begin{table*}[t]
\tbl{Merger classification results by redshift bin. The numbers of galaxies classified as mergers, non-mergers, and unclassified are shown for each redshift range.
The merger fraction (in parentheses) represents the proportion of mergers relative to the total number of galaxies in each bin.}{
\begin{tabular}{crrrr}\hline\hline 
Redshift range & \multicolumn{1}{c}{Merger} & \multicolumn{1}{c}{Non-merger} & \multicolumn{1}{c}{Unclassified} & \multicolumn{1}{c}{Total} \\    \hline
    $0.075\leq z < 0.1$ & 3,794 & 6,806 & 53 & 10,653 ($32.0\pm 0.4$)\\
    $0.1\leq z < 0.15$ & 5,760 & 9,447 & 67 & 15,274 ($45.8\pm 0.4$)\\
    $0.15\leq z < 0.2$ & 3,112 & 4,253 & 28 & 7,393 ($22.2\pm 0.3$)\\ \hline
    All & 12,666 ($38.0\pm 0.4$) & 20,506 ($61.5\pm 0.5$) & 148 ($0.44\pm 0.04$) & 33,320 \\\hline 
\end{tabular}}\label{tab:counts}
\end{table*}

\begin{figure}
 \begin{center}
  \includegraphics[width=0.45\textwidth]{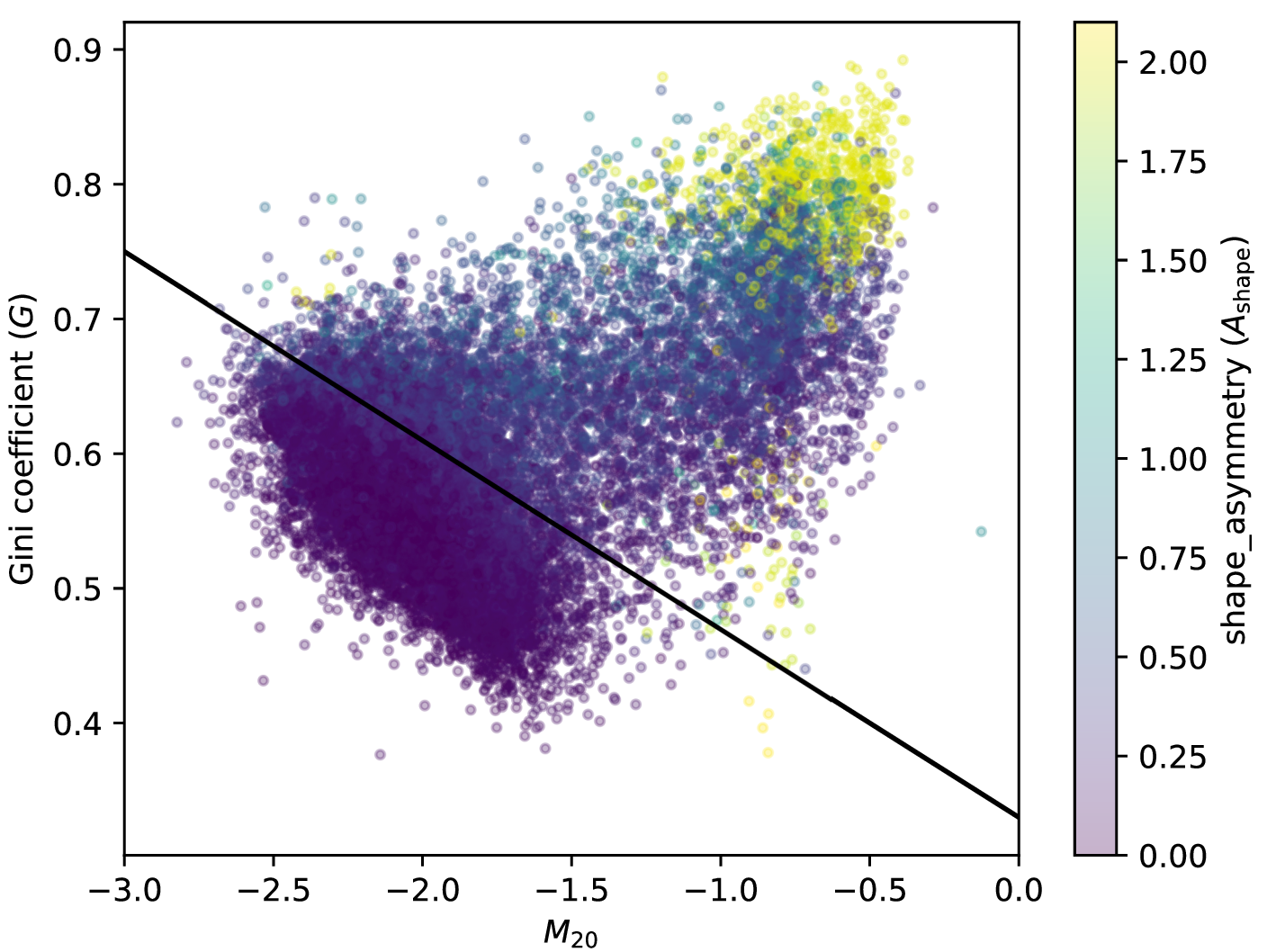}
 \end{center}
\caption{Distribution of galaxies in the Gini--$M_{20}$ plane. The solid line ($G = -0.14\,M_{20} + 0.33$) indicates the merger/non-merger boundary. The color scale represents \texttt{shape\_asymmetry} ($A_{\rm shape}$). Galaxies above the line and with high asymmetry tend to be classified as mergers under our criterion.
{Alt text: A scatter graph, the x axes show the $M_{20}$ from -3.0 to 0.0, the y axes show the Gini coefficient from 0.3 to 0.9.}
}
\label{fig:GM}
\end{figure}

Figure~\ref{fig:GM} shows the distribution of galaxies 
in the Gini--$M_{20}$ plane. The solid line, given by $G = -0.14\,M_{20} + 0.33$, defines the empirical separation boundary, above which galaxies are considered mergers based on Gini--$M_{20}$ statistics alone \citep{Lotz08b}. 
The color scale represents the \texttt{shape\_asymmetry} values.
Most galaxies with high $A_{\mathrm{shape}}$ ($> 0.2$) lie above this line, indicating a strong correlation between the two diagnostics. Notably, some galaxies with low $A_{\mathrm{shape}}$ ($\leq 0.2$) are also found above the boundary, underscoring the need to combine both indicators to avoid false positives and ensure robust classification.

Using the combined criterion ($S(G, M_{20}) > 0$ and $A_{\mathrm{shape}} > 0.2$), we classified 12,666 galaxies (38\% of the full sample) as mergers. 
The full classification breakdown by redshift bin is summarized in table~\ref{tab:counts}.

\begin{figure*}[h]
 \begin{center}
  \includegraphics[width=0.9\textwidth]{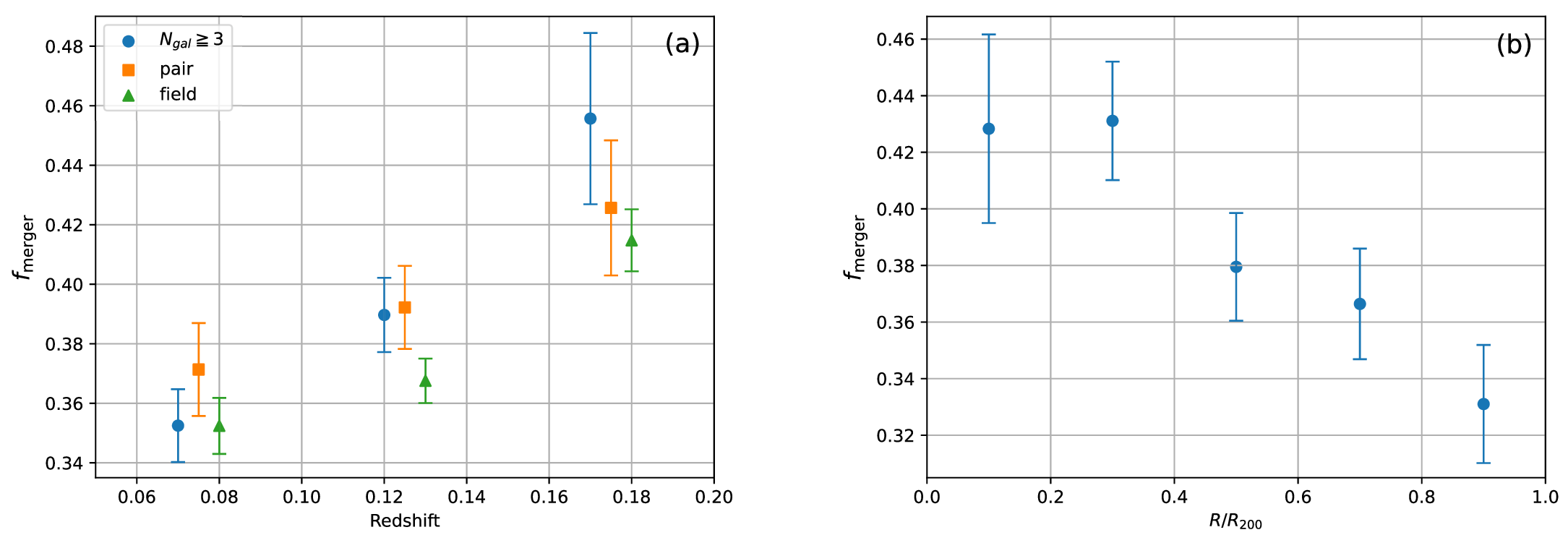}
 \end{center}
 \caption{
Environmental dependence of the merger fraction.
(a) Merger fraction as a function of redshift, shown for cluster member galaxies (blue circle), galaxy pairs (orange square), and field galaxies (green triangle). All subsamples show a positive correlation with redshift.
(b) Merger fraction as a function of normalized cluster-centric distance ($R/R_{200}$). The trend shows a clear increase toward the cluster center, suggesting a density-dependent enhancement in merger activity.
{Alt text: Two graphs. In left panel which labeled a, x axes show the redshift from 0.0 to 0.2, y axes show the merger fraction from 0.34 to 0.48. In right panel which labeled b, x axes show $R/R_{200}$ from 0.0 to 1.0, y axes show the merger fraction 0.3 to 0.46.}
}
\label{fig:results}
\end{figure*}

\subsection{Redshift distribution}
\label{subsec:z}
We investigated how the merger fraction evolves with redshift across three subsamples: cluster member galaxies, galaxy pairs, and field galaxies. Figure~\ref{fig:results}a shows the merger fraction $f_{\mathrm{merger}}$ in three redshift bins: $0.075 \leq z < 0.10$, $0.10 \leq z < 0.15$, and $0.15 \leq z < 0.20$.

All subsamples exhibit an increasing trend in merger fraction with redshift. 
The measured fractions for member galaxies in each redshift bin are $0.352\pm 0.012$, $0.390 \pm 0.012$, and $0.46 \pm 0.03$, respectively. 
Galaxy pairs show similar values: $0.371 \pm 0.016$, $0.392 \pm 0.014$, and $0.43 \pm 0.02$. 
Field galaxies follow with $0.352 \pm 0.009$, $0.368 \pm 0.007$, and $0.415 \pm 0.010$. 
The correlation coefficients for these trends are 0.987, 0.991, and 0.959, respectively.

Although the differences are within statistical uncertainties, member galaxies and galaxy pairs tend to show slightly higher merger fractions than field galaxies. 
This suggests that mergers are more common at higher redshifts and in denser environments, consistent with hierarchical structure formation scenarios.

\subsection{Galaxy cluster radial distribution}
\label{subsec:R}

We further examined the environmental dependence of the merger fraction within galaxy clusters by analyzing its variation with projected radial distance from the cluster center. The radial distance $R$ was normalized by the virial radius $R_{200}$, defined as the radius within which the average density is 200 times the critical density of the Universe. 
Both the cluster center positions and $R_{200}$ values were taken from the group and cluster catalog of \citet{Tempel17}.

Figure~\ref{fig:results}b shows the merger fraction as a function of $R/R_{200}$. The values increase steadily from the outskirts to the central regions: 0.33 $\pm$ 0.02, 0.37 $\pm$ 0.02, 0.38 $\pm$ 0.02, 0.43 $\pm$ 0.02 and 0.43 $\pm$ 0.03, from outer to inner bins. The Spearman's rank correlation coefficient is $-0.886$ (with p-value 0.019), indicating a strong negative radial gradient.
These results imply that mergers are more likely to occur near the cluster center, where the local galaxy density is higher, which is in good agreement with \citet{shibuya25}.
This supports the scenario that dynamical interactions and tidal perturbations are enhanced in dense cluster cores.

\section{Discussion}
\label{sec:discus}
\subsection{Evaluation of classification accuracy}
\label{subsec:class_evaluation}

As the galaxies in our sample do not have associated $P(\mathrm{interaction})$ values, a direct comparison with GALAXY CRUISE classifications is not possible. 
Instead, we compared overall classification trends and performed a visual inspection of selected galaxies.

\begin{figure*}[t]
 \begin{center}
  \includegraphics[width=0.8\textwidth]{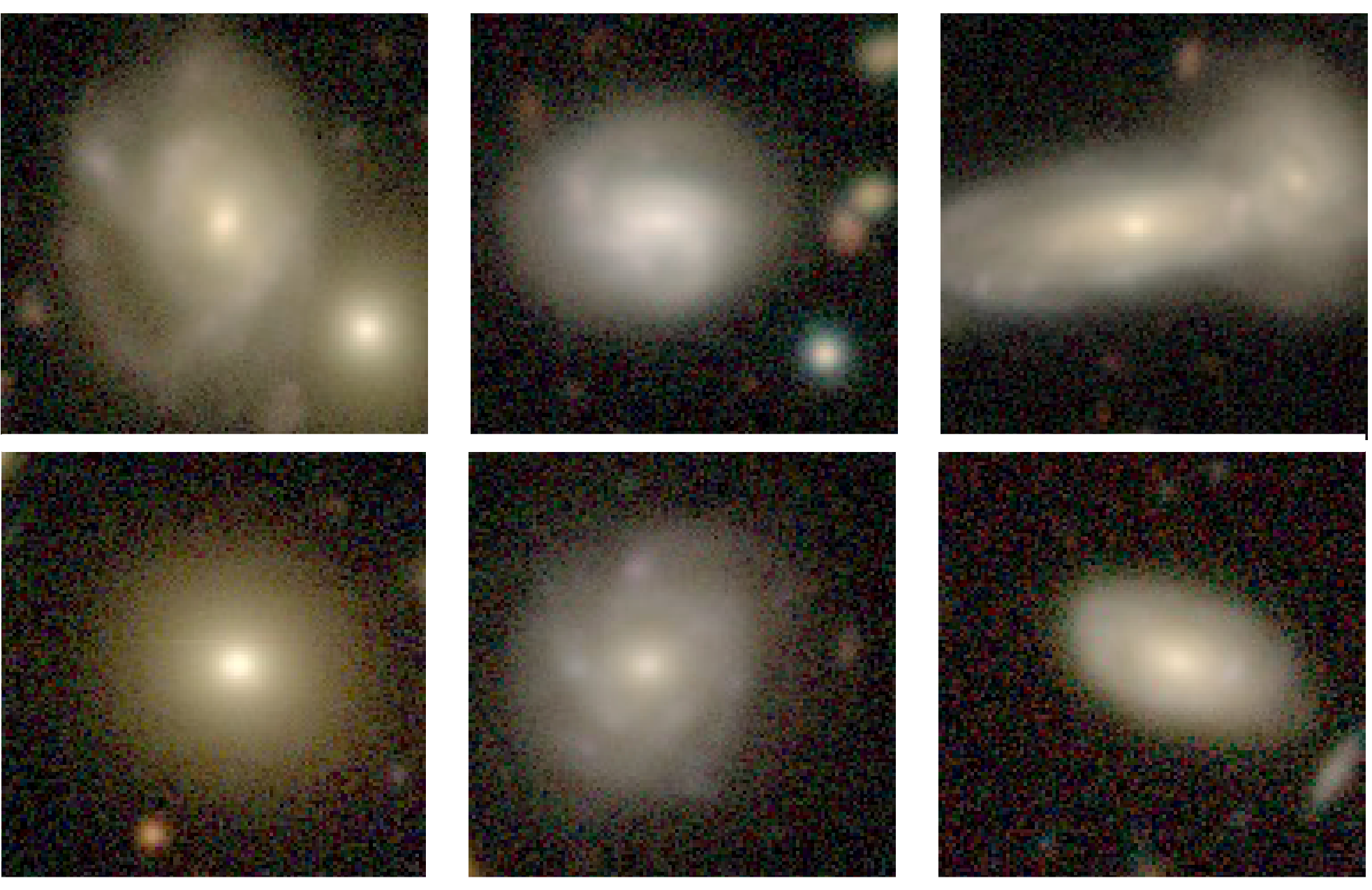}
 \end{center}
 \caption{
 Examples of visual classification results. 
Upper panels show galaxies classified as mergers using our method. 
Lower panels show galaxies classified as non-mergers. Morphological features such as tidal tails and asymmetries are clearly visible in upper panels, supporting the robustness of the classification.
{Alt text: Six images of galaxies. Upper three show mergers. Lower three show non-mergers.}
}\label{fig:check}
\end{figure*}

Table~\ref{tab:comp_GC} summarizes the number of mergers identified using three classification methods: CAS, Gini--$M_{20}$, and our proposed method (equation (\ref{eq:bunrui})). 
The difference in merger fractions between our sample and GALAXY CRUISE is within 3\%, suggesting comparable accuracy.
To further validate our classification, we visually inspected randomly selected samples from both the merger and non-merger categories. Representative examples are shown in figure~\ref{fig:check}. 
The galaxies classified as mergers exhibit clear signs of morphological disturbance, such as asymmetries or tidal features, while the non-merger galaxies appear morphologically regular. 
This supports the overall reliability of our automated classification scheme.

\begin{table}
\tbl{Number of mergers and their fractions identified by different classification methods in this study and in the GALAXY CRUISE dataset. The fractions (in parentheses) are calculated relative to the total sample size in each dataset.}{
\begin{tabular}{crr}
\noalign{\vskip3pt} 
    \hline
    \hline\noalign{\vskip3pt} 
    \multicolumn{1}{c}{Method} & \multicolumn{1}{c}{GALAXY CRUISE} & \multicolumn{1}{c}{This study} \\
    \noalign{\vskip2pt} 
    \hline\noalign{\vskip3pt} 
    CAS & 2,187 ($10.7\pm 0.3$) & 3,077 ($9.23\pm 0.17$) \\
    Gini-$M_{20}$ & 5,868 ($28.7\pm 0.4$) & 8,530 ($25.6\pm 0.3$) \\
    Equation (\ref{eq:bunrui}) & 8,097 ($39.7\pm 0.5$) & 12,666 ($38.0\pm0.4$) \\ 
    \hline\noalign{\vskip3pt} 
\end{tabular}}\label{tab:comp_GC}
\end{table}


\subsection{Redshift dependence of merger fraction}
\label{subsec:discus_z}

Our results confirm a positive correlation between merger fraction and redshift, consistent with previous findings such as those by \citet{Conselice09}. However, our measured merger fractions are approximately three times higher than those reported in earlier studies. This discrepancy may be attributed to several factors:
\begin{itemize}
    \item Differences in redshift range and sample selection: the previous studies focused on galaxies at $0.2 \leq z < 1.2$ observed with e.g., the the Hubble Space Telescope (HST), while our sample lies at $0.075 \leq z < 0.2$ in the HSC-SSP region.
    \item Higher spatial resolution of the HSC data compared to SDSS, allowing for more sensitive detection of morphological disturbances.
    \item Adoption of a new classification scheme combining multiple non-parametric indicators.
\end{itemize}
Taken together, these factors suggest that our methodology enables the detection of a greater number of mergers, particularly those with subtle features that might be missed in previous analyses.

Although our sample spans a relatively narrow redshift range and benefits from the high resolution of HSC imaging, it is important to carefully consider the effect of decreasing angular resolution on physical scales at higher redshifts. This effect is particularly significant for faint and small galaxies, which are more susceptible to resolution loss than their brighter counterparts. If this resolution effect were corrected for, the observed redshift dependence of the merger fraction would likely become even more pronounced.

\subsection{Environmental dependence of merger activity}
\label{subsec:density}

We found that the merger fraction increases toward the centers of galaxy clusters and in higher-density environments. This contrasts with the result from \citet{GC_Tanaka} (see also \cite{Omori}), which suggested a decrease in merger activity in high-density regions.

One possible explanation is the difference in the dominant halo mass scale. Our sample primarily consists of galaxy groups with richness $N < 50$, where relative velocities are lower and mergers can occur more efficiently. 
In contrast, the Tanaka et al. sample likely includes a higher fraction of massive clusters, where high velocity dispersions suppress merger rates. This interpretation is supported by numerical simulations (e.g., \cite{Jian}).

\begin{figure}
 \begin{center}
  \includegraphics[width=0.45\textwidth]{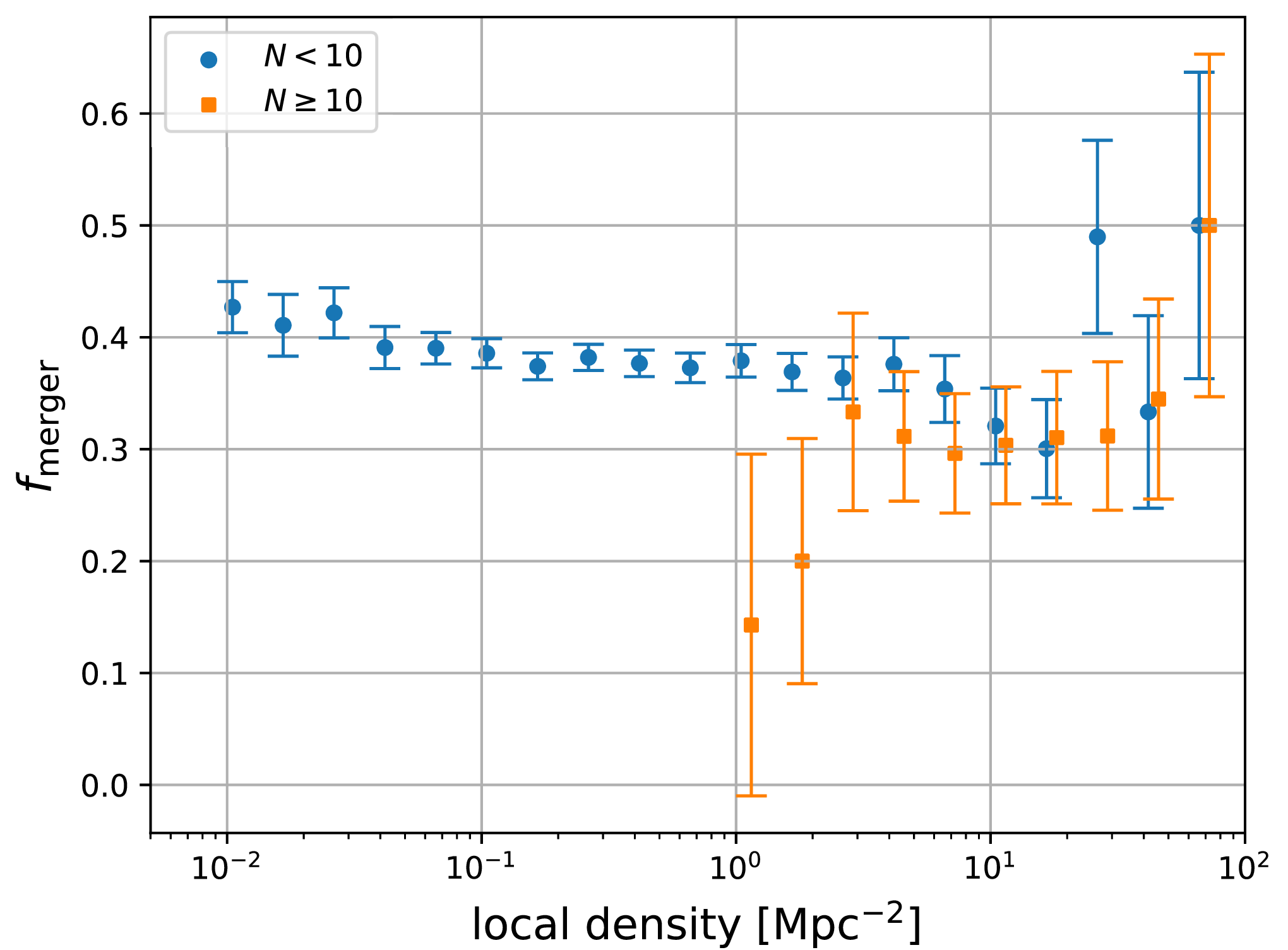}
 \end{center}
\caption{Merger fraction as a function of local galaxy density. 
Blue circles represent systems with richness $N < 10$, and orange squares correspond to $N \geq 10$. The figure shows that denser environments (higher local density) tend to have higher merger fractions, particularly in richer systems.
{Alt text: The x axes show the local galaxy density per square megaparsec from 0.05 to 100 in log scale. The y axes show the merger fraction from -0.05 to 0.7.}
}
\label{fig:ld_N}
\end{figure}

Furthermore, when comparing local density (within to the 5th nearest neighbor) dependence using the same definition as Tanaka et al., as shown in figure~\ref{fig:ld_N}, we find that for galaxies with $N \geq 10$, the merger fraction increases with density, while for $N < 10$, the trend is weaker or ambiguous. 
This suggests that density-enhanced merger activity is most apparent in moderately rich systems. 
The comparison and specific derivation with the local density presented in \citet{GC_Tanaka} (see figure~17 in \cite{GC_Tanaka}) are discussed in Appendix \ref{ap:2}.

\subsection{Connection between mergers and AGN activity}
\label{subsec:AGN}

We explored whether galaxy mergers contribute to AGN activity in galaxy groups and clusters.
AGN were classified based on the BPT diagnostics \citep{BPT}, using the criteria proposed by \citet{Kewley}, \citet{Kauffmann}, and \citet{Schawinski}, as implemented in the SDSS DR10 {\tt emissionLinesPort} table. Galaxies that exceed both the \citet{Kewley} and \citet{Schawinski} thresholds were selected, resulting in a total of 4492 AGN-host galaxies.

\begin{figure}
 \begin{center}
  \includegraphics[width=0.45\textwidth]{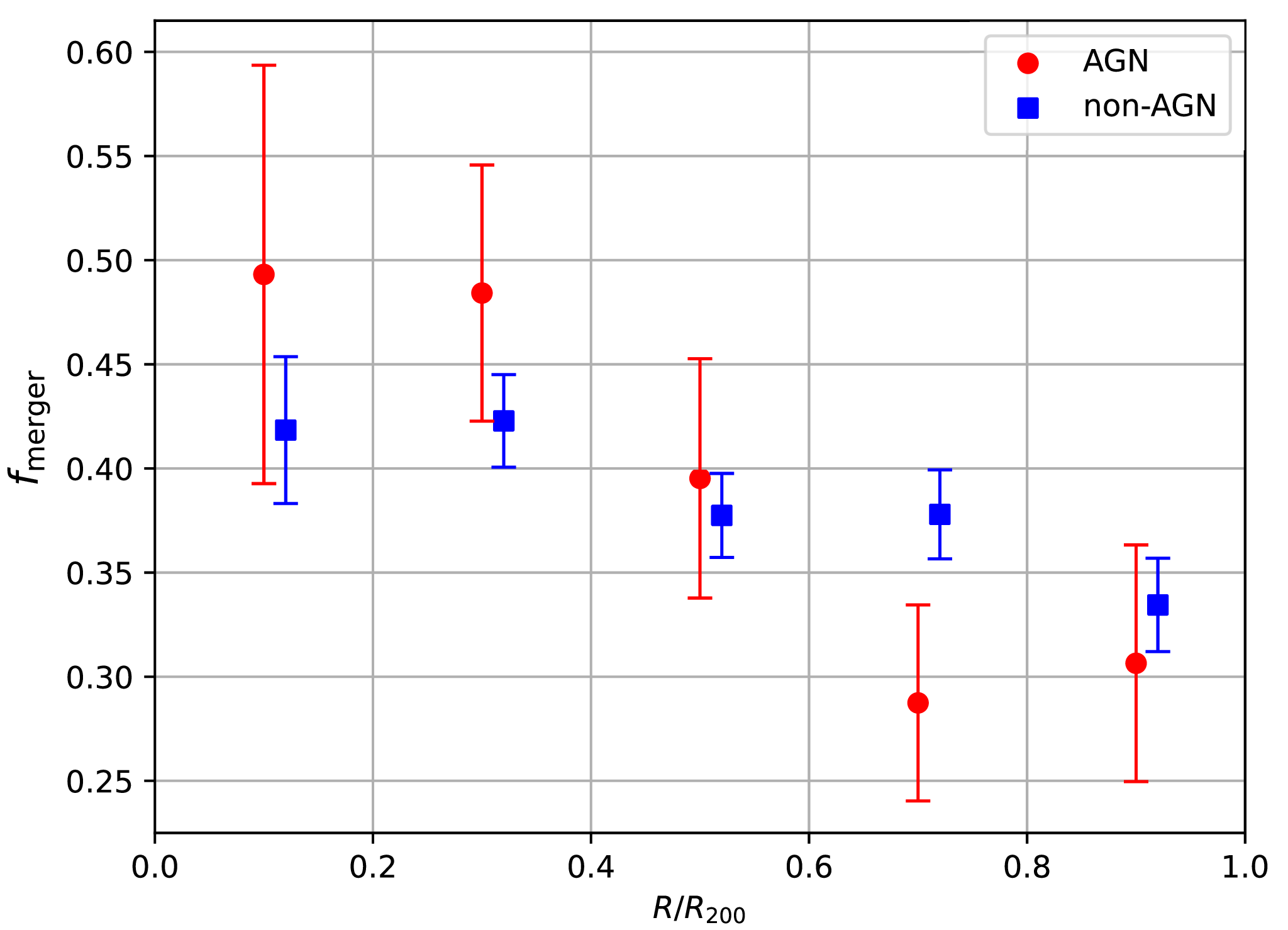}
 \end{center}
\caption{Merger fraction as a function of distance from the cluster center, separated by AGN presence. 
Red circles show galaxies hosting AGNs, and blue squares show non-AGN galaxies. While overall merger fractions are similar, AGN-host galaxies exhibit a mild increase in merger fraction toward the cluster center.
{Alt text: The x axes show the $R/R_{200}$ from 0.0 to 1.0. The y axes show the merger fraction from 0.22 to 0.6.}
}
\label{fig:R_AnA}
\end{figure}

\begin{figure}
 \begin{center}
  \includegraphics[width=0.45\textwidth]{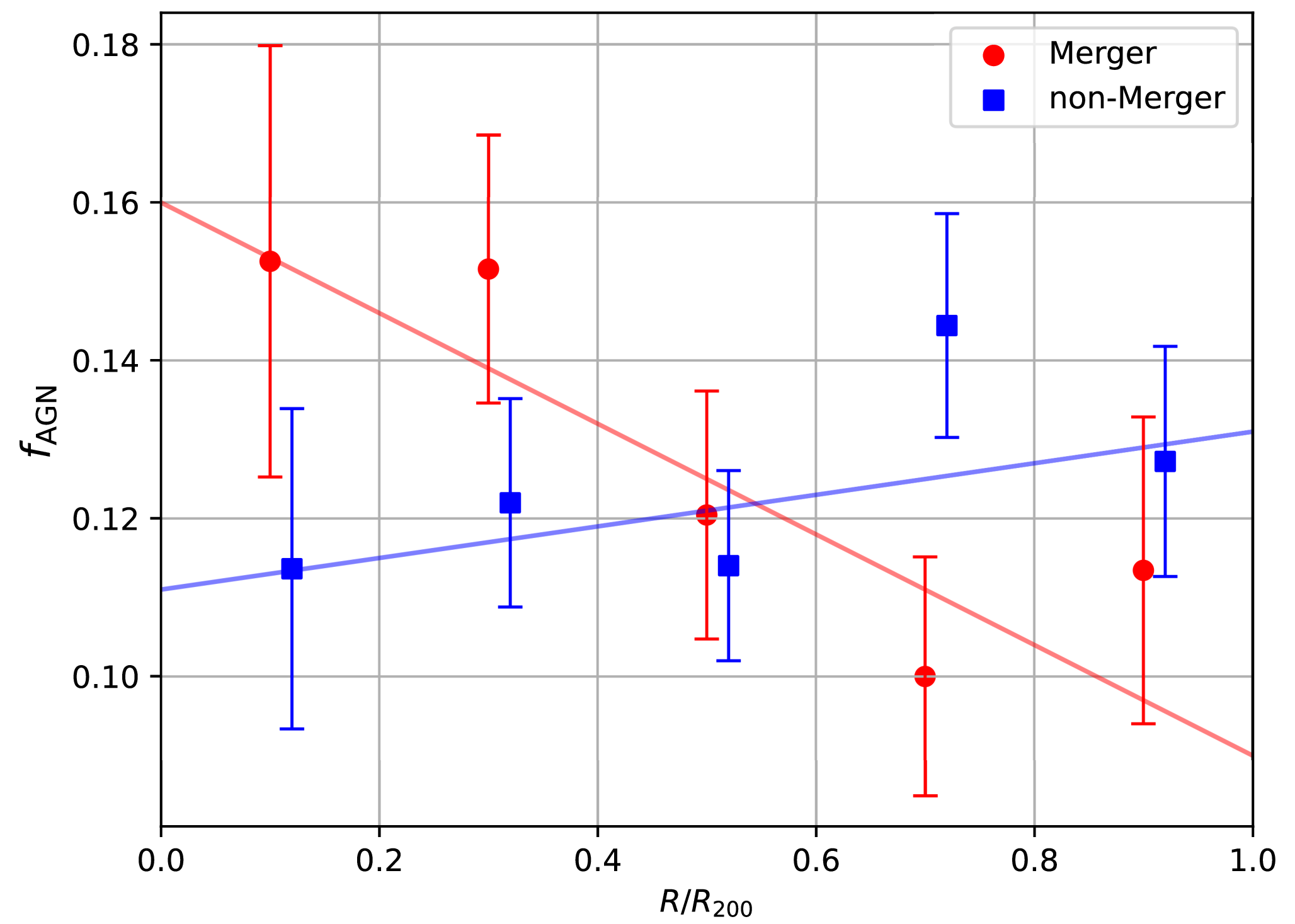}
 \end{center}
\caption{AGN fraction as a function of distance from the center of the galaxy cluster when divided between merger and non-merger galaxies. The red circles in the figure indicate the AGN fractions of the merger galaxies, and the blue squares indicate the AGN fractions of the non-merger galaxies. Straight lines were fitted by the least-squares method with error included, respectively. The value on the horizontal axis is the median of the bins.
{Alt text: The x axes show the $R/R_{200}$ from 0.0 to 1.0. The y axes show the AGN fraction from 0.08 to 0.18.}
}
\label{fig:R_MnM}
\end{figure}

Figure \ref{fig:R_AnA} shows the merger fraction as a function of cluster-centric radius for galaxies with and without hosting AGN.
We observed a slight increase in merger fraction toward cluster centers among AGN-host galaxies, although the overall merger fractions for AGN-host and non-AGN galaxies were statistically consistent.
Conversely, as shown in figure \ref{fig:R_MnM}, when examining AGN fraction ($f_{\rm AGN}$) as a function of cluster-centric distance for mergers and non-mergers separately, we found a mild increasing trend only for merger galaxies. 

We found tentative evidence that galaxy mergers could enhance AGN activity in denser environments, albeit with low statistical significance.
This enhancement of AGN in denser environments has also been found by \citet{Hashiguchi} who reported the AGN fraction shows a significant excess in the cluster center (but see e.g., \cite{Pimbblet} for counterargument).
Recently, \citet{Drigga} also reported that the merger fraction in AGN associated with denser environments is higher than that in field environments in which they used HSC images to classify galaxy mergers.
Those results suggest that galaxy mergers may play an important role in triggering AGN activity in dense environments.
It should be noted, however, that the merger-AGN connection may depend on several factors, including cluster mass (e.g., \cite{Toba24a}), AGN selection methods (e.g., \cite{Galametz}), and AGN luminosity (e.g., \cite{Yuk}).  Moreover, the occurrence of AGN activity also depends on the merger stage (e.g., \cite{Toba22b, Bickley24b, Ellison25}), which should be kept in mind in interpreting these results.



\section{Summary}
\label{sec:summary}
In this study, we investigated the environmental dependence of galaxy mergers using high-resolution imaging data from the Hyper Suprime-Cam Subaru Strategic Program (HSC-SSP). Our main findings are summarized as follows:
\begin{itemize}
    \item We developed a new morphological classification method that combines Gini--$M_{20}$ statistics with the \texttt{shape\_asymmetry} parameter, enhancing the detection of both centrally concentrated and morphologically disturbed galaxies.
    
    \item Applying this method to a sample of 33,320 galaxies at $0.075 \leq z < 0.2$, we identified 12,666 mergers, corresponding to a merger fraction of 38\%. Our classification results show good agreement with GALAXY CRUISE visual annotations.
    
    \item The merger fraction increases with redshift for all subsamples (member galaxies, galaxy pairs, and field galaxies), consistent with hierarchical formation scenarios.
    
    \item A clear radial dependence was observed in clusters, with merger fractions increasing toward the cluster center. This trend supports enhanced merger activity in denser environments, especially within galaxy groups.
    
    \item We also found tentative evidence that mergers may contribute to AGN triggering in high-density regions, although the statistical significance is limited.
\end{itemize}
These results demonstrate the effectiveness of combining non-parametric diagnostics for large-scale merger identification and highlight the importance of both cosmic time and environment in galaxy interaction processes.

\begin{ack}
The Hyper Suprime-Cam (HSC) collaboration includes the astronomical communities of Japan and Taiwan, and Princeton University. 
The HSC instrumentation and software were developed by the National Astronomical Observatory of Japan (NAOJ), the Kavli Institute for the Physics and Mathematics of the Universe (Kavli IPMU), the University of Tokyo, the High Energy Accelerator Research Organization (KEK), the Academia Sinica Institute for Astronomy and Astrophysics in Taiwan (ASIAA), and Princeton University.  
Funding was contributed by the FIRST program from the Japanese Cabinet Office, the Ministry of Education, Culture, Sports, Science and Technology (MEXT), the Japan Society for the Promotion of Science (JSPS), Japan Science and Technology Agency  (JST), the Toray Science  Foundation, NAOJ, Kavli IPMU, KEK, ASIAA, and Princeton University.

This paper is based on data collected at the Subaru Telescope and retrieved from the HSC data archive system, which is operated by Subaru Telescope and Astronomy Data Center (ADC) at NAOJ. Data analysis was in part carried out with the cooperation of Center for Computational Astrophysics (CfCA) at NAOJ.  We are honored and grateful for the opportunity of observing the Universe from Maunakea, which has the cultural, historical and natural significance in Hawaii.

This paper makes use of software developed for Vera C. Rubin Observatory. We thank the Rubin Observatory for making their code available as free software at \url{http://pipelines.lsst.io/}. 

The Pan-STARRS1 Surveys (PS1) and the PS1 public science archive have been made possible through contributions by the Institute for Astronomy, the University of Hawaii, the Pan-STARRS Project Office, the Max Planck Society and its participating institutes, the Max Planck Institute for Astronomy, Heidelberg, and the Max Planck Institute for Extraterrestrial Physics, Garching, The Johns Hopkins University, Durham University, the University of Edinburgh, the Queen’s University Belfast, the Harvard-Smithsonian Center for Astrophysics, the Las Cumbres Observatory Global Telescope Network Incorporated, the National Central University of Taiwan, the Space Telescope Science Institute, the National Aeronautics and Space Administration under grant No. NNX08AR22G issued through the Planetary Science Division of the NASA Science Mission Directorate, the National Science Foundation grant No. AST-1238877, the University of Maryland, Eotvos Lorand University (ELTE), the Los Alamos National Laboratory, and the Gordon and Betty Moore Foundation.
Data analysis was in part carried out on the Multi-wavelength Data Analysis System operated by the Astronomy Data Center (ADC) and the Large-scale data analysis system co-operated by the Astronomy Data Center and Subaru Telescope, National Astronomical Observatory of Japan.

This work is supported by Japan Science and Technology Agency (JST) Support for Pioneering Research Initiated by the Next Generation (SPRING),  Grant Number JPMJSP2115.
This work is also supported by the Japan Society for the Promotion of Science (JSPS) KAKENHI Grant Numbers 20K04027 (NO), 23K22537 (YT), 21K03632, and 25K07359 (MI).
NO acknowledges partial support by the Organization for the Promotion of Gender Equality at Nara Women's University.
\end{ack}



\appendix\label{appendix} 
\section{Adjustments to statmorph}\label{ap:1}

In its default configuration, \texttt{statmorph} allows the user to input a mask image as a two-dimensional array of the same size as the science image. However, the masking information in the HSC images used in this study was insufficient to fully exclude neighboring sources. To address this, we modified the pipeline to focus only on the central object in each image, masking all other sources.

In addition, we revised the object segmentation process to improve the accuracy of morphological measurements. Specifically, we adjusted the segmentation map threshold used in the calculation of Gini and $M_{20}$ statistics. This adjustment incorporates the \texttt{shape\_asymmetry} value as a control variable.Originally, \texttt{shape\_asymmetry} was computed after the Gini--$M_{20}$ statistics. In our modified version, it is calculated immediately beforehand so that it can be used to dynamically set the segmentation threshold. 

We updated the segmentation criterion equation (\ref{eq:a1}) to equation (\ref{eq:a2}).
Segmentation maps generated using equation (\ref{eq:a1}) tend to be too small for merger galaxies and too large for non-mergers, 
which can compromise morphological accuracy. This is because equation (\ref{eq:a1}) selects pixels above \texttt{ellip\_annulus\_mean\_flux} for inclusion in the segmentation area. The smaller this flux threshold is, the fainter the structure that can be captured—important for detecting tidal features in mergers. Therefore, a lower threshold is desirable for mergers, while a higher threshold is more appropriate for non-mergers.

We found that \texttt{shape\_asymmetry} is an effective parameter for adjusting the segmentation threshold because mergers tend to exhibit higher \texttt{shape\_asymmetry} values than non-mergers. The new formulation in equation (\ref{eq:a2}) improves robustness to background noise and enables more reliable morphological diagnostics, particularly in faint or low-surface-brightness galaxies.

\begin{align}\label{eq:a1}
\texttt{cutout\_smooth >= ellip\_annulus\_mean\_flux}, 
\end{align}
\begin{align}\label{eq:a2}
\texttt{cutout\_smooth} >= \texttt{ellip\_annulus\_mean\_flux} \cdot \frac{0.2}{\texttt{asym} + 0.25}.
\end{align}

\section{Estimation of Local Galaxy Density}\label{ap:2}

To facilitate comparison with figure~17 in \citet{GC_Tanaka}, we calculated local galaxy density using a similar method. The procedure was as follows:
\begin{enumerate}
    \item Compute the line-of-sight velocity ($v_{\rm rad}$) from the spectroscopic redshift.
    \item Select galaxies within $\pm$1000 km/s of the target galaxy in radial velocity space.
    \item Compute the angular distance to the 5th nearest galaxy in this projected sample.
    \item Convert this angular distance to physical distance ($r_5$) using the angular diameter distance.
    \item Calculate the local surface density as $\Sigma = \frac{5}{\pi r_5^2}$ (in units of galaxies per Mpc$^2$).
\end{enumerate}
Note that this density estimate does not account for masked regions or incompleteness due to observational effects, and should therefore be interpreted as a simplified approximation.

\begin{figure}
 \begin{center}
  \includegraphics[width=0.45\textwidth]{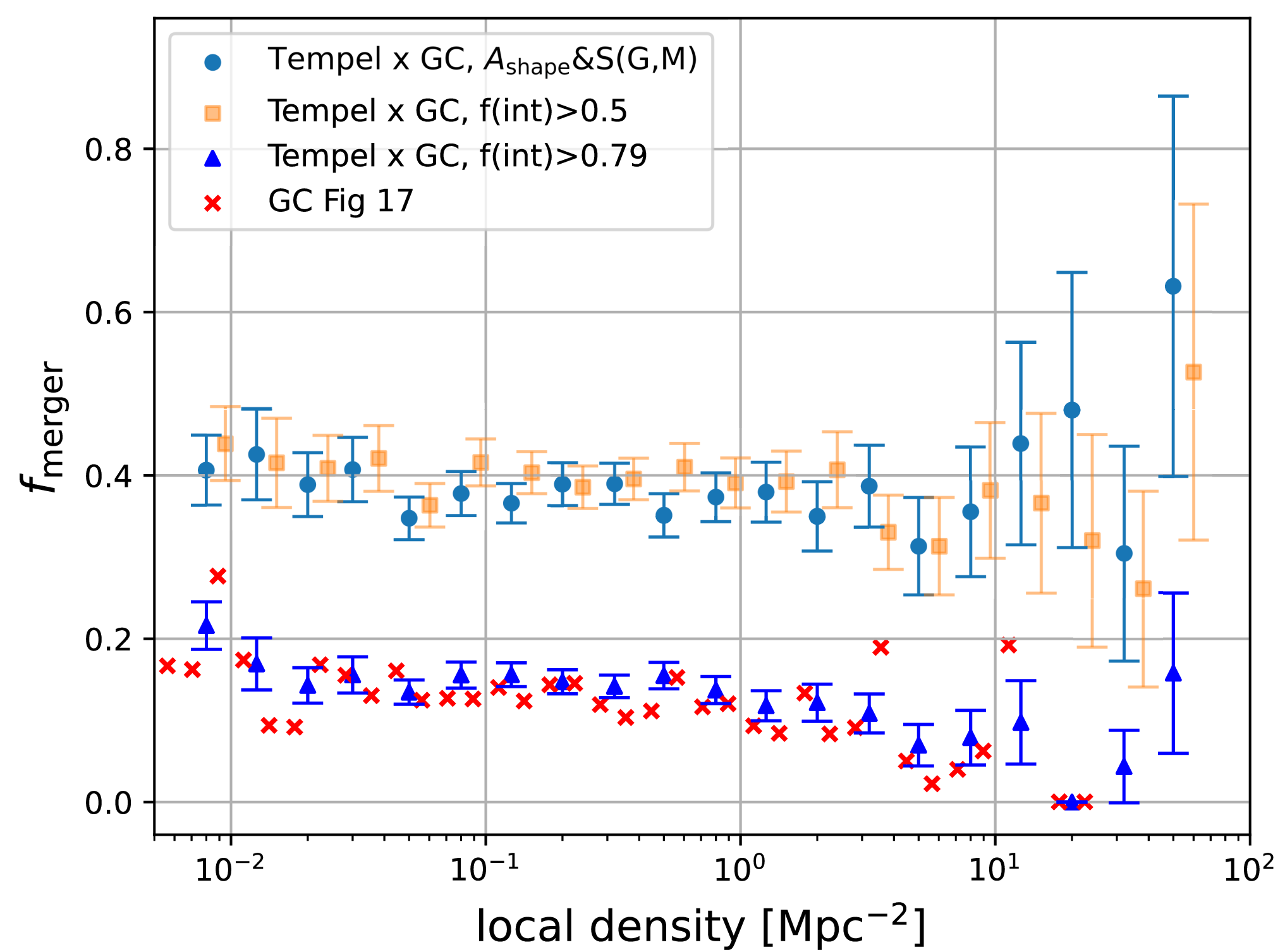}
 \end{center}
\caption{Comparison of merger fractions against local density. Each adopts a different classification criterion for the sample after cross-matching to classify the merger. Light blue circle data points are the classification criterion of this study, blue triangle data points are the GALAXY CRUISE criteria, and orange square data points are the less stringent GALAXY CRUISE criteria. Red crosses are the values taken from fig.~17 (right panel) in \citet{GC_Tanaka}.
{Alt text: The x axes show the local galaxy density per square megaparsec from 0.05 to 100 in log scale. The y axes show the merger fraction from -0.05 to 0.95.}
}
\label{fig:ld_comp}
\end{figure}

To compare classification performance under different merger identification schemes, we cross-matched our sample with the GALAXY CRUISE sample. Figure~\ref{fig:ld_comp} shows the merger fraction as a function of local density under three different criteria:
\begin{itemize}
    \item Our method (light blue circles)
    \item GALAXY CRUISE with $P(\mathrm{interaction}) > 0.79$ (blue triangles)
    \item GALAXY CRUISE with relaxed threshold $P(\mathrm{interaction}) > 0.5$ (orange squares)
\end{itemize}

For reference, we also plot the values reported in \citet{GC_Tanaka} (red crosses). Our method yields a higher merger fraction than the stricter GALAXY CRUISE threshold, and is broadly consistent with the relaxed version. This suggests that our approach is sensitive to intermediate merger candidates that may be missed by stricter visual thresholds.

\bibliographystyle{pasj}
\bibliography{cite}

\end{document}